\documentclass[aip,amsmath,amssymb,reprint]{revtex4-1}

\usepackage[utf8]{inputenc}
\usepackage[T1]{fontenc}
\usepackage[english]{babel}
\usepackage[section]{placeins}
\usepackage{mathptmx}
\usepackage{etoolbox}
\usepackage{dcolumn}
\usepackage{bm}
\usepackage{xcolor}
\usepackage{float}
\usepackage{physics}
\usepackage{siunitx}
\usepackage{placeins}
\usepackage{soul} 

\usepackage{algorithm}
\usepackage{algpseudocode}

\usepackage{hyperref}
\hypersetup{
    colorlinks=true,
    citecolor=black,
    linkcolor=black,
    filecolor=black,
    urlcolor=black,
}

\usepackage{cleveref}
\Crefname{equation}{Eq.}{Eqs.}
\Crefname{figure}{Fig.}{Figs.}
\Crefname{section}{Sec.}{Secs.}

\usepackage{caption}
\usepackage{subcaption}
\usepackage{graphicx}
\graphicspath{{./}}

\makeatletter
\def\@email#1#2{%
 \endgroup
 \patchcmd{\titleblock@produce}
  {\frontmatter@RRAPformat}
  {\frontmatter@RRAPformat{\produce@RRAP{*#1\href{mailto:#2}{#2}}}\frontmatter@RRAPformat}
  {}{}
}%
\makeatother

\begin{document}

\preprint{AIP/123-QED}

\title{Gas-generating reactive flows in bicontinuous catalyst support structures}

%

\author{J.M.P. Beunen}
\affiliation{Helmholtz Institute Erlangen-Nürnberg für Erneuerbare Energien (IET--2), Cauerstr.~1, 91058 Erlangen, Germany}

\author{J. Harting}
\email{j.harting@fz-juelich.de}
\affiliation{Helmholtz Institute Erlangen-Nürnberg für Erneuerbare Energien (IET--2), Cauerstr.~1, 91058 Erlangen, Germany}%
\affiliation{Department of Chemical and Biological Engineering and Department of Physics, Friedrich-Alexander-Universit\"at Erlangen-N\"urnberg, Cauerstr.~1, 91058 Erlangen, Germany}

\date{\today}

\begin{abstract}
A major challenge in the field of heterogeneous catalysis is selecting an optimal catalyst support structure. Commercially available structures can be easily manufactured at scale, but their stochastic nature makes their chemical and transport properties suboptimal. This is particularly relevant for gas-generation reactions, where non-uniformity of a porous structure leads to bubble trapping. Such trapping impedes the flow of reactants to catalyst sites, leading to conversion inefficiencies. Previous experimental work demonstrated that spinodally-derived architectures, in particular bicontinuous interfacially jammed emulsion gels (bijels), can alleviate these issues and deliver superior performance. However, to the best of our knowledge, numerical studies to optimize the operating conditions for such a morphology have not been performed yet. In this work, we aim to close this gap using color-gradient lattice Boltzmann simulations of reactive flows with a novel central moments collision operator. We develop an analytical model to predict catalyst performance based on our simulation data. Our findings show that this type of morphology can achieve very high conversion efficiencies. Moreover, we demonstrate that its catalyst performance can be optimized using superhydrophilic surface coatings.
\end{abstract}

\maketitle

\section{Introduction}\label{sec:introduction}

Gas-generating reactive flows are omnipresent in industrial applications involving heterogeneous catalysis. Prominent examples include petroleum cracking, large-scale inorganic chemical production, and energy conversion \& storage \cite{Thomas2015PrinciplesPracticeHeterogeneous}. These applications generally involve a solid catalyst, combined with reactants and products in liquid and/or gaseous phases \cite{Ertl2008HandbookHeterogeneousCatalysis}. In most applications, porous structures are used as catalyst supports due to their very high surface-to-volume ratio, which enables efficient use of the catalyst material. However, in the case of gas-generating reactive flows, blocking of catalyst sites by the generated gas can occur. This reduces the rate of the chemical reaction and, therefore, the efficiency of the often very expensive catalyst material \cite{Solymosi2022NucleationRatedeterminingStep}. Unfortunately, this issue is worsened by the stochastic nature of commercially available catalyst support structures. These structures generally have non-uniform pore sizes and a significant number of pore necks, leading to hold-up and trapping of reaction product within the structure \cite{Kolitcheff2017TortuosityMesoporousAlumina}.

Given the limitations of stochastic support structures, significant research efforts have been dedicated to obtaining more optimized morphologies in recent years \cite{Aguirre2020OpencellFoamsCatalysts,Balzarotti2021PeriodicOpenCellular,Li2024BioinspiredMonolithicCatalyst,Eckendorfer2024PeriodicOpenCellular}. One feasible solution to obtain more efficient morphologies is by making use of structures derived from spinodally decomposing liquids \cite{Stratford2005ColloidalJammingInterfaces,Beunen2026PerformanceOptimizationBijels,Sultan2025OpenPoresParticles}. In particular, adding nanoparticles that arrest the spinodal decomposition of the liquids to obtain bicontinuous interfacially jammed emulsion gels (bijels) seems to be a viable pathway to obtain highly optimized morphologies \cite{Stratford2005ColloidalJammingInterfaces,Herzig2007BicontinuousEmulsionsStabilized}.

Bijel formation can occur when particles with partial wettability are present in a mixture of two spinodally decomposing liquids. The free energy of the total system is reduced when the particles remain in contact with both fluids, enabling their trapping at the interface \cite{Stratford2005ColloidalJammingInterfaces,Cates2008BijelsNewClass}. These trapping energies can be several orders of magnitude larger than the thermal energy, leading to permanent capture of the particles at the interface. As the interface coarsens, the amount of available surface area for the particles steadily decreases. This forces the particles together to the point that they arrest the interfacial coarsening \cite{Stratford2005ColloidalJammingInterfaces,Cates2008BijelsNewClass,White2008InfluenceParticleComposition}. At this point, two cases can be distinguished. Either droplets covered with particles have been formed, known as a Pickering-Ramsden emulsion \cite{Ramsden1904SeparationSolidsSurfacelayers,Pickering1907CXCVIEmulsions}. Or, a quasi-two-dimensional sheet of particles percolating through the entire domain, known as a bijel.

The possibility for the formation of bijels was first shown by means of computer simulations in 2005 by Stratford \textit{et al.} \cite{Stratford2005ColloidalJammingInterfaces}. Two years later, their formation was experimentally demonstrated by Herzig \textit{et al.} \cite{Herzig2007BicontinuousEmulsionsStabilized}. Since this pioneering work, bijels have been experimentally realized using various fluids and particle types \cite{Cates2008BijelsNewClass, White2011InversionParticlestabilizedEmulsions, Tavacoli2011NovelRobustVersatile, Witt2013BijelReinforcementDroplet, Firoozmand2015FoodgradeBijelsBased, Hijnen2015BijelsStabilizedUsing, Haase2016SituMechanicalTesting, Cai2017BijelsFormedDirect, Huang2017BicontinuousStructuredLiquids, DiVitantonio2021FabricationApplicationBicontinuous, Pizzetti2021BiphasicPorousStructures, Ching2021RapidProductionBicontinuous, Sprockel2023FabricationBijelsSubmicron, Amirfattahi2024FabricationBijelsSolvent}. Furthermore, several simulation studies have explored their design space and properties \cite{Kim2010BijelsContainingMagnetic, Jansen2011BijelsPickeringEmulsions, Frijters2012EffectsNanoparticlesSurfactant, Gunther2013LatticeBoltzmannSimulations,
    Frijters2014SelfassembledPorousMedia, Carmack2015NumericalSimulationsBijel, Carmack2018TuningThinfilmBijels, Bonaccorso2020ShearDynamicsConfined, Steenhoff2025AnalysisBijelFormation}.

After solidification of one of the two fluid phases in a bijel, the resulting structure exhibits properties that are highly beneficial for heterogeneous catalysis applications. These structures possess a high flow permeability, uniform pore sizes, a fully connected pore space, and no dead ends \cite{Clegg2020BijelsBicontinuousParticlestabilized}. Therefore, droplets and bubbles can flow through these porous structures without significant changes in Laplace pressure, thereby reducing the chance for bubble trapping. This makes these structures an ideal candidate for highly optimized support structures.

The usability of bijel-derived morphologies for chemically reactive flows was first experimentally demonstrated by Witt \textit{et al.}~\cite{Witt2016MicrostructuralTunabilityCocontinuous}. In that study, the authors constructed electrodes from bijel-derived porous structures, reporting a fifty percent improvement in energy density over previously tested electrode morphologies. The superior performance of these electrodes was further confirmed by Gross \textit{et al.}~\cite{Gross2021MitigatingBubbleTraffic}, who studied bijel-derived electrodes in the context of hydrogen gas generation by water electrolysis. Their findings suggest significantly faster product gas expulsion for bijel-derived electrodes and a lower overpotential requirement, further illustrating the superior morphology. Manufacturability, thermal stability, and robustness of bijel-derived porous structures were demonstrated by Allasia \textit{et al.}~\cite{Allasia2024BijelbasedMesophotoreactorIntegrated}. In their research, they studied the photodegradation of methylene blue, further emphasizing the potential for usage in industrial settings. Given these promising experimental findings, we investigated the optimization of bijel-derived morphologies for miscible reactive flows for a range of parameters in our previous work \cite{Beunen2026PerformanceOptimizationBijels}. Here, we found that bijel-derived support structures offer superior performance for miscible reactive flows, yielding up to more than $2.7$ times the output of similar stochastic structures. However, to the best of our knowledge, numerical studies on the potential of these morphologies for chemically reactive immiscible flows are still missing. Therefore, in this paper, we numerically study the optimization possibilities of a bijel-derived catalyst support structure for chemically reactive immiscible flows. 

The remainder of this article is structured as follows: first, we discuss the numerical method to simulate chemically reactive immiscible flows in \Cref{sec:numerical-method}. Hereafter, we describe the morphological properties of the investigated catalyst supports and how they have been generated in \Cref{sec:porous-structures}. In \Cref{sec:chemically-reactive-flows}, we describe the mapping of experimental studies to the simulations we conduct. The results of these simulations are elaborated upon in \Cref{sec:results}. A conclusion to the paper is presented in \Cref{sec:conclusion}. Finally, we report on a novel formulation of the simulation algorithm in Appendix \ref{sec:generalized-central-moments-collision-operator}, where we write the central moments collision operator in a generalized form for arbitrary equilibria.

\section{Numerical method}\label{sec:numerical-method}

\subsection{Multiphase model}
We use the color-gradient lattice Boltzmann method to simulate reactive flows through porous structures. This well-established algorithm allows for accurate simulation of multiphase flows through porous structures, combined with excellent parallelization on massively parallel architectures \cite{Benzi1992LatticeBoltzmannEquation,Succi2001LatticeBoltzmannEquation,Kruger2017LatticeBoltzmannMethod}. The algorithm is based on a lattice discretization of the Boltzmann equation in physical space, velocity space, and time. The lattice vectors are denoted with $\boldsymbol{c}_i$ and have the values stored in the $i$th column of the matrix
\begin{multline}
    \mathbf{M} = \left[
        \begin{array}{rrrrrrrrrrrr}
            1 & -1 & 0 & 0  & 0 & 0  & 1 & 1  & 1 & 1  & -1 & -1 \\
            0 & 0  & 1 & -1 & 0 & 0  & 1 & -1 & 0 & 0  & 1  & -1 \\
            0 & 0  & 0 & 0  & 1 & -1 & 0 & 0  & 1 & -1 & 0  & 0
        \end{array}\right.
        \\
        \left.
        \begin{array}{rrrrrrr}
            -1 & -1 & 0 & 0  & 0  & 0  & 0 \\
            0  & 0  & 1 & 1  & -1 & -1 & 0 \\
            1  & -1 & 1 & -1 & 0  & 0  & 0
        \end{array}
        \right].
\end{multline}
In addition, their weights are denoted as $w_i$, with the weight function given by
\begin{equation}
    w\left(\left|\boldsymbol{c}_i\right|^2\right) = \begin{cases}
                                                        1 / 3 & \left|\boldsymbol{c}_i\right|^2 = 0 \\
                                                        1 / 18 & \left|\boldsymbol{c}_i\right|^2 = 1 \\
                                                        1 / 36 & \left|\boldsymbol{c}_i\right|^2 = 2
    \end{cases}.
    \label{eq:d3q19_weight_function}
\end{equation}
This yields a velocity space discretization in 19 distinct vectors (D3Q19)\cite{Qian1992LatticeBGKModels}. Furthermore, we use the general convention for the lattice spacing and time step $\Delta x = \Delta t = 1$. This gives the default expression for the lattice speed of sound $c_s = 1 / \sqrt{3}$. We consider two immiscible fluid components, the reactant $r$ and the product $p$, which both evolve according to their own distribution functions $f_i^r$ and $f_i^p$. For one of these components, $k = r$ or $p$, the local mass and momentum can be computed as
\begin{equation}
    \rho^k = \rho_{0} \sum_i f_i^k \quad \text{and} \quad \rho^k \boldsymbol{u}^k = \rho_{0} \sum_i f_i^k \boldsymbol{c}_i
    \label{eq:moments_components}
\end{equation}
from these distribution functions. Here, $\rho_{0} = 1$ is the unit mass for both fluids. Additionally, the local mixture mass and local mixture momentum can be computed as
\begin{equation}
    \rho^\mathrm{m} = \sum_k \rho^k \quad \text{and} \quad \rho^\mathrm{m} \boldsymbol{u}^\mathrm{m} = \sum_k \rho^k \boldsymbol{u}^k.
    \label{eq:moments_mixture}
\end{equation}
Each fluid component $k$ obeys the lattice Boltzmann equation
\begin{equation}
    f_i^k\left(\boldsymbol{x}+\boldsymbol{c}_i \Delta t, t+\Delta t\right)=f_i^k(\boldsymbol{x}, t)+\Omega_i^k(\boldsymbol{x}, t)
    \label{eq:lattice_boltzmann_equation}
\end{equation}
at a lattice site with position $\boldsymbol{x}$ at time $t$. The collision operator $\Omega_i^k(\boldsymbol{x}, t)$ is split into a combination of three separate sub-operators
\begin{equation}
    \Omega_i^k = \left(\Omega_i^k\right)^{\left(3\right)}\left[\left(\Omega_i^k\right)^{\left(1\right)} + \left(\Omega_i^k\right)^{\left(2\right)}\right],
    \label{eq:operator_decomposition}
\end{equation}
with $\left(\Omega_i^k\right)^{\left(1\right)}$ being the perturbation operator, $\left(\Omega_i^k\right)^{\left(2\right)}$ being the single-phase collision operator, and $\left(\Omega_i^k\right)^{\left(3\right)}$ being the recoloring operator. These sub-operators will be further elaborated upon below.

First, the contribution of the perturbation operator is added to the distribution functions. This operator is applied to exert forces on the fluid components, and implemented according to the exact difference scheme proposed by Kupershtokh \textit{et al.} \cite{Kupershtokh2009EquationsStateLattice}. We opt for the exact difference scheme for two reasons. First, it can be written in a way that is completely independent of the operations performed by the single-phase collision operator. As a result, $\left(\Omega_i^k\right)^{\left(2\right)}$ can be written as if the algorithm were force-free, enabling full decoupling between the perturbation operator and the single-phase collision operator. Second, it is fully indifferentiable. As a result, it is possible to apply this scheme to the distribution functions of every component separately without violating consistent hydrodynamics \cite{Asinari2008ConsistentLatticeBoltzmann}. The operator can be written as
\begin{equation}
    \left(\Omega_i^k\right)^{\left(1\right)} = f_i^{\mathrm{eq}}\left(\rho^k, \boldsymbol{u}^\mathrm{m} + \Delta\boldsymbol{u}^k\right) - f_i^{\mathrm{eq}}\left(\rho^k, \boldsymbol{u}^\mathrm{m}\right),
    \label{eq:operator_perturbation}
\end{equation}
where we use the standard second-order truncated equilibrium distribution function
\begin{equation}
    f_i^{\mathrm{eq}}\left(\rho, \boldsymbol{u}\right) = w_i \rho\left[1+\frac{\boldsymbol{u} \cdot \boldsymbol{c}_i}{c_s^2}+\frac{\left(\boldsymbol{u} \cdot \boldsymbol{c}_i\right)^2}{2 c_s^4}-\frac{\boldsymbol{u} \cdot \boldsymbol{u}}{2 c_s^2}\right],
    \label{eq:equilibrium_distribution}
\end{equation}
where $\rho$ and $\boldsymbol{u}$ correspond to the according densities and velocities. Additionally, we define the velocity shift in \Cref{eq:operator_perturbation} as
\begin{equation}
    \Delta\boldsymbol{u}^k = \frac{\mathbf{F}^k}{\rho^k} \Delta t.
    \label{eq:delta_u}
\end{equation}
The force $\mathbf{F}^k$ is computed for every fluid component separately, and includes a surface tension and an external force contribution. The surface tension contribution is curvature-dependent and based upon the continuum surface force concept \cite{Brackbill1992ContinuumMethodModeling, Lishchuk2003LatticeBoltzmannAlgorithm}. As a result,
\begin{equation}
    \mathbf{F}^k = \left(\frac{1}{2} \sigma H \nabla \rho^N\right) \frac{\rho^k}{\rho^\mathrm{m}} + \rho^k \mathbf{a}^k,
    \label{eq:continuum_surface_force}
\end{equation}
with $\sigma$ being the surface tension between reactant and product. The acceleration vector $\mathbf{a}^k$ can be set to a non-zero value when there is an external body force on any of the fluids. Moreover, $H$ is the mean interface curvature between both fluids, which can be calculated as
\begin{equation}
    H=-[(\mathbf{I}-\mathbf{n} \otimes \mathbf{n}) \cdot \nabla] \cdot \mathbf{n},
    \label{eq:interface_curvature}
\end{equation}
with $\mathbf{I}$ being the identity matrix, and
\begin{equation}
    \mathbf{n} = - \frac{\nabla \rho^N}{\left|\nabla \rho^N\right|}
    \label{eq:color_gradient_normal}
\end{equation}
being the normalized gradient of the color function
\begin{equation}
    \rho^N\left(\boldsymbol{x}, t\right) = \frac{\rho^r\left(\boldsymbol{x}, t\right) - \rho^p\left(\boldsymbol{x}, t\right)}{\rho^r\left(\boldsymbol{x}, t\right) + \rho^p\left(\boldsymbol{x}, t\right)}.
    \label{eq:color_function}
\end{equation}
The partial derivatives in \Cref{eq:continuum_surface_force}, \Cref{eq:interface_curvature}, and \Cref{eq:color_gradient_normal} are approximated using an isotropic finite difference scheme. Using the structure of the lattice, we compute the partial derivatives as
\begin{equation}
    \nabla \phi\left(\boldsymbol{x}, t\right) = \frac{1}{c_s^2} \sum_i w_i \boldsymbol{c}_i \phi\left(\boldsymbol{x}+\boldsymbol{c}_i \Delta t, t\right),
    \label{eq:finite_difference_scheme}
\end{equation}
for any of the required variables $\phi$.

Second, the single-phase collision operator is applied to the distribution function of the fluid mixture, often called the color-blind distribution function, $f_i^\mathrm{m} = f_i^r + f_i^p$. This operator is used to reconstruct the behavior predicted by the Navier-Stokes equations at the level of the fluid mixture. It relaxes the color-blind distribution function to the equilibrium distribution for the fluid mixture computed using $\rho^\mathrm{m}$ and $\boldsymbol{u}^\mathrm{m}$ as input for \Cref{eq:equilibrium_distribution}. We can write the single-phase collision operator as
\begin{equation}
    \left(\Omega_i^k\right)^{\left(2\right)} = \frac{\rho^k}{\rho^\mathrm{m}} \Omega_i^{\mathrm{CM}}.
    \label{eq:single_phase_collision}
\end{equation}
Here, $\Omega_i^{\mathrm{CM}}$ is a reformulation of the central-moment-based collision operator proposed by De Rosis and Coreixas \cite{DeRosis2020MultiphysicsFlowSimulations}. This reformulation enables the usage of an identical algorithm for an entire class of equilibrium distribution functions, using only five instead of nineteen central moments. In Appendix \ref{sec:generalized-central-moments-collision-operator}, we further elaborate on how to implement this generalized and more efficient form of the operator. Besides, the operator contains only a single relaxation time $\tau$, which sets the kinematic viscosity
\begin{equation}
    \nu = c_s^2 \left(\tau - \frac{1}{2}\right)
    \label{eq:kinematic_viscosity}
\end{equation}
in an equivalent way to the BGK collision operator. However, given the drastic stability gain caused by relaxing central moments, we are able to perform simulations with kinematic viscosities that are two orders of magnitude lower than what can be achieved with the BGK collision operator \cite{Coreixas2019ComprehensiveComparisonCollision}.

Third and finally, recoloring operators are applied to enforce immiscibility of both fluid components. While the perturbation operator imposes a surface tension between the fluid components, immiscibility is not guaranteed. To this end, we apply the recoloring operators \cite{Latva-Kokko2005DiffusionPropertiesGradientbased}
\begin{equation}
    \begin{aligned}
        \left(\Omega_i^r\right)^{(3)} &= f_i^r +\beta \frac{\rho^r \rho^p}{\rho^\mathrm{m}} w_i \frac{\boldsymbol{c}_i \cdot \mathbf{n}}{\left|\boldsymbol{c}_i\right|},\\
        \left(\Omega_i^p\right)^{(3)} &= f_i^p -\beta \frac{\rho^r \rho^p}{\rho_
        \mathrm{m}} w_i \frac{\boldsymbol{c}_i \cdot \mathbf{n}}{\left|\boldsymbol{c}_i\right|}.
    \end{aligned}
    \label{eq:operator_recoloring}
\end{equation}
Here, $\beta$ is a parameter that controls the thickness of the numerical interface. To reduce spurious currents to a minimum, we keep this parameter constant in the simulations at $\beta = 0.7$.

\subsection{Boundary conditions}
In all the simulations, periodic boundary conditions are applied to the outer faces of the simulation box. Additionally, we apply the halfway bounce-back boundary condition
\begin{equation}
    f_i^k\left(\boldsymbol{x}, t+\Delta t\right) = f_{\bar{i}}^k(\boldsymbol{x}, t)
    \label{eq:bounce_back}
\end{equation}
on all solid boundaries to enforce a no-slip condition at fluid-solid interfaces. Here, $\bar{i}$ denotes the index of the lattice vector pointed in the opposite direction of $i$ and into a solid node. Additionally, at fluid nodes adjacent to a fluid-solid interface, we make use of the wetting scheme proposed by Akai \textit{et al.} \cite{Akai2018WettingBoundaryCondition}. To this end, we estimate the color function at the solid sites by means of a lattice-weighted average
\begin{equation}
    \rho^N\left(\boldsymbol{x}, t\right) = \frac{\sum_i w_i \rho^N\left(\boldsymbol{x} + \boldsymbol{c}_i \Delta t, t\right)\: l\left(\boldsymbol{x} + \boldsymbol{c}_i \Delta t\right)}{\sum_i w_i \: l\left(\boldsymbol{x} + \boldsymbol{c}_i \Delta t\right)},
    \label{eq:color_lattice_weighted_average}
\end{equation}
with $l\left(\boldsymbol{x}\right)$ being an indicator function for fluid nodes, evaluating to $1$ for fluid nodes and $0$ for solid nodes. Hereafter, we compute the normalized gradient of the color function in \Cref{eq:color_gradient_normal}, as well as two new realigned normal vectors
\begin{equation}
    \mathbf{n}_\pm = \left(\cos\pm\theta - \frac{\sin\pm\theta\cos\theta^\prime}{\sin\theta^\prime}\right) \mathbf{n}_s + \frac{\sin\pm\theta}{\sin\theta^\prime} \mathbf{n}
    \label{eq:realigned_normal_vectors}
\end{equation}
for the fluid sites where at least one lattice vector is pointing into a solid node. In this equation, $\theta^\prime = \arccos\left(\mathbf{n}_s \cdot \mathbf{n}\right)$ and $\theta$ is the contact angle that we impose. We evaluate the Euclidean distances between $\mathbf{n}_{\pm}$ and $\mathbf{n}$ at every lattice site, then $\mathbf{n}$ is replaced with either $\mathbf{n}_{+}$ or $\mathbf{n}_{-}$ depending on which vector has the smallest Euclidean distance to $\mathbf{n}$. The normal vector of the solid boundary $\mathbf{n}_s$ at these sites is evaluated by the eighth-order isotropic discretization \cite{Xu2017LatticeBoltzmannSimulation}
\begin{equation}
    \mathbf{n}_s\left(\boldsymbol{x}\right) = \frac{\sum_i \omega_i\left(\left|\boldsymbol{e}_i\right|^2\right) s\left(\boldsymbol{x} + \boldsymbol{e}_i \Delta t\right)\boldsymbol{e}_i}{\left|\sum_i \omega_i\left(\left|\boldsymbol{e}_i\right|^2\right) s\left(\boldsymbol{x} + \boldsymbol{e}_i \Delta t\right)\boldsymbol{e}_i\right|},
    \label{eq:normal_vector_solid}
\end{equation}
with $s\left(\boldsymbol{x}\right) = 1 - l\left(\boldsymbol{x}\right)$ being an indicator function for solid nodes. The eighth-order weight function is given by \cite{Sbragaglia2007GeneralizedLatticeBoltzmann}
\begin{equation}
    \omega\left(\left|\boldsymbol{e}_i\right|^2\right) = \begin{cases}
                                                             4 / 45 & \left|\boldsymbol{e}_i\right|^2 = 1 \\
                                                             1 / 21 & \left|\boldsymbol{e}_i\right|^2 = 2 \\
                                                             2 / 105 & \left|\boldsymbol{e}_i\right|^2 = 3 \\
                                                             5 / 504 & \left|\boldsymbol{e}_i\right|^2 = 4 \\
                                                             1 / 315 & \left|\boldsymbol{e}_i\right|^2 = 5 \\
                                                             1 / 630 & \left|\boldsymbol{e}_i\right|^2 = 6 \\
                                                             1 / 5040 & \left|\boldsymbol{e}_i\right|^2 = 8
    \end{cases}.
    \label{eq:eigth_order_weight_function}
\end{equation}
In line with previous papers, we select the eighth-order weight function in order to reduce the spurious currents close to the contact lines \cite{Xu2017LatticeBoltzmannSimulation, Li2021ModelingThreephaseDisplacement}.

Furthermore, we use a reactive boundary condition that converts reactant into product at several fluid sites neighboring the solid sites. For this purpose, we compute the number of nearest neighboring solid sites at a fluid site as
\begin{equation}
    \text{NN}\left(\boldsymbol{x}\right) = l\left(\boldsymbol{x}\right) \sum_{i : \left|\boldsymbol{c}_i\right|^2 = 1} s\left(\boldsymbol{x} + \boldsymbol{c}_i \Delta t\right),
    \label{eq:nearest_solid_neighbors}
\end{equation}
which we then use to modify the fluid densities according to first-order kinetics as
\begin{equation}
    \frac{\partial \rho^r\left(\boldsymbol{x}, t\right)}{\partial t} = - \xi \: \rho^r\left(\boldsymbol{x}, t\right) r\left(\boldsymbol{x}\right) \text{NN}\left(\boldsymbol{x}\right),
    \label{eq:reactive_boundary_condition_reactant}
\end{equation}
and
\begin{equation}
    \frac{\partial \rho^p\left(\boldsymbol{x}, t\right)}{\partial t} = + \xi \: \rho^r\left(\boldsymbol{x}, t\right) r\left(\boldsymbol{x}\right) \text{NN}\left(\boldsymbol{x}\right),
    \label{eq:reactive_boundary_condition_product}
\end{equation}
for the reactant and the product, respectively. Here, the $\xi$ variable denotes the reaction rate. The density modifications in these equations are implemented on a population level by directly subtracting and adding values from the populations at these sites in a uniform manner. Additionally, $r\left(\boldsymbol{x}\right)$ is an indicator function being 1 for sites where the reaction is activated and 0 where it is not. The selection of sites where the reaction is activated is geometry-dependent. Therefore, its exact formulation will be explained later in \Cref{sec:chemically-reactive-flows}.

\section{Porous structures}\label{sec:porous-structures}
To generate bijel-derived porous structures, we make use of algorithms described in our previous papers \cite{Jansen2011BijelsPickeringEmulsions, Frijters2012EffectsNanoparticlesSurfactant, Beunen2026PerformanceOptimizationBijels}. Since we want to focus specifically on the influence of morphology on reactive immiscible flows, we consider only a single bijel-derived porous geometry. In particular, we opt for a large-scale porous geometry that we investigated in our previous work \cite{Beunen2026PerformanceOptimizationBijels}, with a particle volume fraction $\Xi = 0.20$, a fluid-fluid ratio of $5/9$, and an equilibrium particle contact angle of $103$ degrees. Hereafter, we subsample this geometry from $1024^3$ to $256^3$ lattice sites using the procedure described in the original paper. Additionally, we remove solid sites floating around in the void space and void sites in the solid space, which are small artifacts of the subsampling procedure.

\begin{figure}
    \centering
    \includegraphics[width=\columnwidth,trim=50 50 50 100, clip]{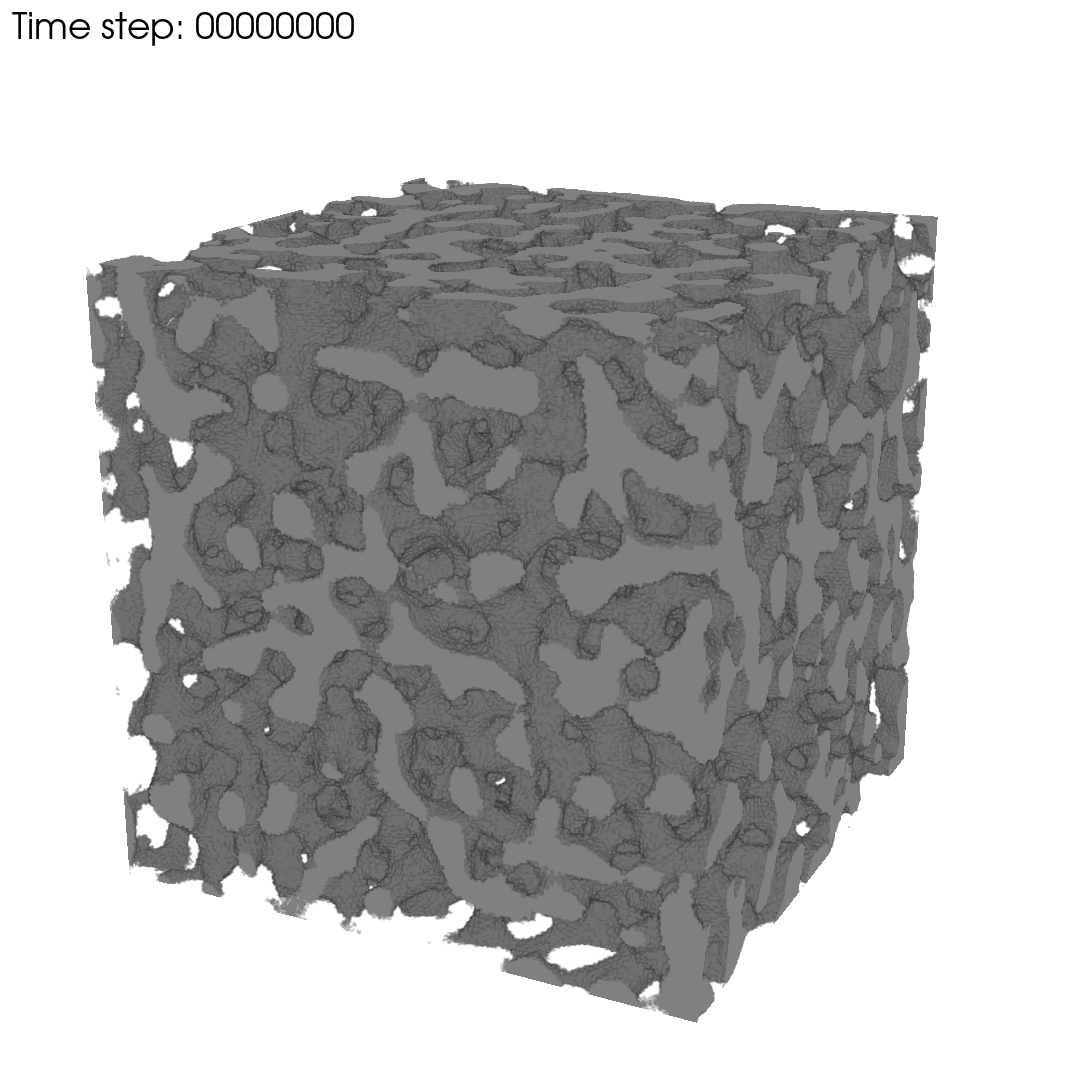}
    \caption{A visualization of the bijel-derived porous structure used in the simulations. The pore structure is rendered in gray, with the void space being transparent.}
    \label{fig:rock_visualization3d}
\end{figure}

A visualization of the resulting bijel-derived porous morphology can be found in \Cref{fig:rock_visualization3d}. This particular structure is chosen as a model geometry for two reasons. First, we know that these parameter combinations yield proper bijels, with no intermediate structures that also exhibit features of Pickering emulsions. Second, the average pore radius $R \approx 10 \Delta x$ that we obtain from the subsampled geometry yields pore sizes that are large enough to accurately simulate effects related to Laplace pressure \cite{Leclaire2017GeneralizedThreedimensionalLattice}. At the same time, this pore size is also not so large that computational costs would become prohibitive.

Next, we analyze the morphological properties of the bijel-derived porous structure. In \Cref{tab:porous_structure_properties}, several morphological properties that can be directly computed from the geometry are further elaborated upon. The surface area of the structure is computed as
\begin{equation}
    A = \sum_{\boldsymbol{x}} \text{NN}\left(\boldsymbol{x}\right)
    \label{eq:surface_area}
\end{equation}
by making use of \Cref{eq:nearest_solid_neighbors}. The porosity is defined as
\begin{equation}
    \phi = \frac{\sum_{\boldsymbol{x}} l\left(\boldsymbol{x}\right)}{\sum_{\boldsymbol{x}} 1}.
    \label{eq:porosity}
\end{equation}
Furthermore, we display values computed based on the permeabilities of the porous structure in \Cref{tab:porous_structure_properties}. The permeabilities are computed by making use of the procedures outlined by Narvaez \textit{et al.}~\cite{Narvaez2010EvaluationPressureBoundary, Narvaez2010QuantitativeAnalysisNumerical}. As a first step, we rewrite Darcy's law to gain an expression for the permeability
\begin{equation}
    \kappa = \frac{\nu \left<u\right>}{a},
    \label{eq:permeability}
\end{equation}
with $\left<u\right>$ the average fluid velocity in the direction of an externally applied body force. This body force is set by applying an acceleration on the fluid, where the length of the acceleration vector, i.e., the acceleration constant, is denoted by $a$. Hereafter, we perform single-component lattice Boltzmann simulations to compute the permeability of the structure in the three Cartesian directions. For these  simulations, we set the length $a = 10^{-6}$ and $\nu = 0.1$. As a result, the Reynolds numbers are much smaller than 1, ensuring that we are in the Stokes flow regime.

As postulated in our previous work, we hypothesize that a large part of the efficiency of any porous structure as a chemical reactor is determined by the permeability as well as the available reactive surface area. Therefore, we also show the value of the reactor efficiency number
\cite{Beunen2026PerformanceOptimizationBijels}
\begin{equation}
    \eta = \kappa \left(\frac{A}{V}\right)^2
\end{equation}
in \Cref{tab:porous_structure_properties}. This allows for a straightforward comparison between the bijel-derived porous structure and others concerning their suitability as a chemical reactor.

\begin{table}
    \centering
    \begin{tabular}{|l|l|}
        \hline
        \textbf{Property}                                        & \textbf{Value}                        \\
        \hline
        Surface area $A$                                         & $\approx 2.892 \cdot 10^6 \Delta x^2$ \\
        Porosity $\phi$                                          & $\approx 0.604$                       \\
        Volume $V$                                               & $= 256^3 \Delta x^3$                  \\
        Surface-area-to-volume ratio $A/V$                       & $\approx 0.172 \: \Delta x^{-1}$      \\
        Average pore radius $R$                                  & $\approx 9.109 \Delta x$              \\
        Averaged single-phase permeability $\left<\kappa\right>$ & $\approx 6.132 \Delta x^2$            \\
        Reactor efficiency number $\eta$                         & $\approx 0.182$                       \\
        \hline
    \end{tabular}
    \caption{A table with relevant geometrical properties of the bijel-derived porous structure that are directly related to the morphology. All the values are computed in lattice units.}
    \label{tab:porous_structure_properties}
\end{table}

\section{Chemically reactive flows}\label{sec:chemically-reactive-flows}

\subsection{Dimensionless numbers}\label{sec:dimensionless-numbers}
To perform realistic simulations of immiscible chemically reactive flows, we distinguish three dimensionless numbers. These are the Bond number, an alternative Bond number based on a pressure gradient over the porous structure, and the Damköhler number. In particular, we want to map our simulations to previous experimental results of gas-generating reactions in bijel-derived porous structures that use water as a solvent \cite{Gross2021MitigatingBubbleTraffic, Allasia2024BijelbasedMesophotoreactorIntegrated}. Therefore, we make use of experimental parameters of water at Standard Ambient Temperature and Pressure (SATP) in \Cref{tab:experimental_parameters}. To distinguish between simulation variables and variables in experiments, we will write experimental variables in SI units with a tilde on top, like $\widetilde{\rho}$ for the density.

\begin{table}
    \centering
    \begin{tabular}{|l|l|}
        \hline
        \textbf{Property}                 & \textbf{Value}                                          \\
        \hline
        Density $\widetilde{\rho_w}$            & $\approx 997.0 \: \si{\kilo\gram\per\meter\cubed}$      \\
        Dynamic viscosity $\widetilde{\mu_w}$   & $\approx 0.890 \:\si{\milli\pascal\second}$             \\
        Kinematic viscosity $\widetilde{\nu_w}$ & $\approx 0.893 \: \si{\milli\meter\squared\per\second}$ \\
        Surface tension $\widetilde{\sigma_w}$  & $\approx 0.0721 \: \si{\newton\per\meter}$              \\
        \hline
    \end{tabular}
    \caption{Relevant properties of water at Standard Ambient Temperature and Pressure (SATP) \cite{Haynes2017CRCHandbookChemistry}.}
    \label{tab:experimental_parameters}
\end{table}

To estimate realistic values for the Bond number
\begin{equation}
    \mathrm{Bo} = \frac{\Delta\rho g R^2}{\sigma},
    \label{eq:bond_number}
\end{equation}
we need to estimate values for the average pore radius in experiments. We can infer realistic values from scanning electron microscopy images of bijel-derived porous structures described in Gross \textit{et al.} \cite{Gross2021MitigatingBubbleTraffic}. This gives experimental pore radii in the range of several microns, $\widetilde{R} \approx 5 \: \si{\micro\meter} - 25 \: \si{\micro\meter}$. Additionally, we need to estimate the density difference $\Delta\rho$ between the reactant and product fluid. Here, we can assume that the product fluid has negligible weight relative to the reactant, i.e., $\widetilde{\Delta\rho} \approx \widetilde{\rho_w}$. In particular, for a hydrogen generation reaction as investigated by Gross \textit{et al.}, the hydrogen density is $\widetilde{\rho_h} \approx 0.0807 \: \si{\kilo\gram\per\meter\cubed}$ under SATP conditions \cite{Haynes2017CRCHandbookChemistry}. This density barely increases due to Laplace pressure; for the smallest pore radius, we obtain a pressure differential $\widetilde{\Delta P} = 2 \widetilde{\sigma_w} / \widetilde{R} \approx 0.280 \: \si{\bar}$. Finally, to estimate the acceleration constant $g$ on the fluids, we use the maximum value for buoyancy-driven flow, which is the gravitational acceleration $\widetilde{g} \approx 9.81 \si{\meter\per\second\squared}$. In the absence of pressure gradients over the structure, this yields an experimental Bond number $\widetilde{\mathrm{Bo}} < 10^{-4}$. In other words, buoyancy-related effects can be considered negligible for this particular setup.

Second, we define a Bond number based on a pressure gradient over the porous structure as
\begin{equation}
    \mathrm{Bo}_p = \frac{\nabla p R^2}{\sigma}.
    \label{eq:bond_number_pressure}
\end{equation}
In this equation, $\nabla p$ is an external pressure gradient. For simulations, we write the pressure gradient as $\nabla p = \rho a$, where $\rho$ is the density of the carrier fluid, and $a$ is an acceleration constant. As a result, the pressure gradient can be implemented as a body force. While a relatively wide range of pressure gradients can be applied depending on the experiment, we make use of a moderate pressure gradient that can be realistically achieved without compromising on the integrity of the porous structure. For this purpose, we set $\mathrm{Bo}_p = 0.01$ in the simulations. We can then compute the pressure gradient in experiments using \Cref{eq:bond_number_pressure}, the experimental pore radius $\widetilde{R}$, and the properties of water in \Cref{tab:experimental_parameters}. This yields experimental pressure gradients of $\widetilde{\nabla p} \approx 12 \: \si{\bar\per\meter} - 288 \: \si{\bar\per\meter}$, where the exact value depends on the pore radius in experiments, $\widetilde{R}$.

Third, we vary the Damköhler number in the simulations. Given that the Damköhler number depends on the catalyst loading, the observed values in experiments can span several orders of magnitude. To compute the Damköhler number in the simulations, we need to define both a flow timescale and a chemical timescale. For the flow timescale, we first define the volumetric flow rate through the porous structure as
\begin{equation}
    Q = \frac{\kappa V a}{\nu L},
    \label{eq:volumetric_flow_rate}
\end{equation}
with $V$ being the volume occupied by the structure, and $L$ the reactor length. Hereafter, we can compute a prediction of the residence time, and hence the flow timescale, as
\begin{equation}
    t_{f} = \frac{V}{Q} = \frac{\nu L}{\kappa a}.
    \label{eq:flow_timescale}
\end{equation}
For the chemical timescale, we use the fact that the chemical reaction is modelled according to first-order kinetics. Hence, the reactivity scales linearly with the reactive surface area $A_r$, and we can define a chemical timescale as
\begin{equation}
    t_{\xi} = \frac{V}{\xi A_r}.
    \label{eq:reactive_timescale}
\end{equation}
These two timescales can then be combined to define the Damköhler number
\begin{equation}
    \mathrm{Da} = \frac{t_{f}}{t_{\xi}} = \frac{\xi A_r \nu L}{\kappa V a}.
    \label{eq:damkohler_number}
\end{equation}
The Damköhler number can be varied in experiments and simulations by changing the catalyst loading or by applying a pressure gradient over the sample. Given that this number can be set to basically any value via the catalyst loading, we investigate the range where both timescales are comparable in this paper, i.e., $\mathrm{Da} = 0.1 - 10.0$.

\subsection{Numerical parameter mapping}
To obtain the correct dimensionless numbers as computed for the experiments in the previous section, we set a total of five parameters in the simulations. These are the kinematic viscosity $\nu$, the acceleration constant $a$, the surface tension $\sigma$, the reaction rate $\xi$, and the reactive surface area $A_r$ via the indicator function $r\left(\boldsymbol{x}\right)$. Of these five parameters, four appear only in \Cref{eq:damkohler_number}. However, \Cref{eq:bond_number} and \Cref{eq:damkohler_number} are coupled via the acceleration constant $a$.

To reduce the computational requirements of the simulations, we would like to reduce the number of time steps, thereby maximizing reactivity within the structures. For this purpose, we set the reaction rate to its maximum possible value, $\xi = 1.0$, enabling instant conversion of reactant to product at reactive sites. Furthermore, we compute the reactive surface area as
\begin{equation}
    A_r = \sum_{\boldsymbol{x}} r\left(\boldsymbol{x}\right) \text{NN}\left(\boldsymbol{x}\right),
    \label{eq:surface_area_reactive}
\end{equation}
which is dependent on the particular form of the indicator function $r\left(\boldsymbol{x}\right)$. While one might trivially be inclined to use $r\left(\boldsymbol{x}\right) = 1$ everywhere, this leads to incorrect physics. Instead of bubble generation, a film of product fluid is formed at the solid-fluid interface. To alleviate this issue, we make use of bubble nucleation sites instead. Given the thickness of the numerical fluid interface, we need many sites inbetween bubble nucleation sites to avoid immediate coalescence. Hence, for every $64^3$ block of lattice sites, we select only a single site with one neighboring rock node as a bubble nucleation site. Therefore, in line with \Cref{eq:surface_area_reactive}, only on such a site $r\left(\boldsymbol{x}\right) = 1$, while all the other lattice sites in the $64^3$ block have $r\left(\boldsymbol{x}\right) = 0$. This yields a quasi-random distribution of bubble nucleation sites in the porous structure, with a total reactive surface area $A_r = 64$. Moreover, we keep the reaction rate and form of $r\left(\boldsymbol{x}\right)$ constant across all simulations, which can be interpreted as using the same catalyst loading in all runs.

Given that the experiments are heavily surface tension dominated, we set the surface tension in the simulations to the maximum possible value $\sigma = 0.3$, right up to the stability limit. This increases the maximum possible acceleration constants that we can set for \Cref{eq:bond_number} and \Cref{eq:bond_number_pressure}. Since the fluids move faster with higher acceleration constants, this decreases the duration of the simulations. Besides, since the Bond number $\mathrm{Bo}$ in \Cref{eq:bond_number} is practically zero, buoyancy-related effects are negligible. As a result, the acceleration constants for both fluids are determined by pressure gradients only, and we just set the acceleration constant $a$. Filling in the values for the simulations in \Cref{eq:bond_number_pressure} gives an acceleration constant of $a \approx 3.616 \cdot 10^{-5} \Delta x / \Delta t^2$. Given that the pressure gradient acts on both fluids, we apply this acceleration to both fluids via the acceleration vector $\mathbf{a}^k$ in \Cref{eq:continuum_surface_force}. Specifically, we apply an acceleration along the first Cartesian axis, i.e., $\mathbf{a}^k = \left(\begin{array}{ccc}
a & 0 & 0 \end{array}\right)^{\top}$. Hereafter, the Damköhler number can be set by varying the kinematic viscosity $\nu$. Given the stability properties of central moments collision operators, the relaxation time $\tau$ can be made very small without running into numerical stability problems. Hence, given \Cref{eq:kinematic_viscosity}, the kinematic viscosity can be varied over a large range as well, enabling us to vary the Damköhler number over the two orders of magnitude that we want to study.

\section{Results}\label{sec:results}

\subsection{Theoretical model}\label{sec:theoretical-model}
To study the behavior of immiscible reactive flows in the simulations, we first examine the system theoretically. For this purpose, we define the total masses of the reactant and product in the simulation box as
\begin{equation}
    m^r\left(t\right) = \sum_{\boldsymbol{x}} \rho^r\left(\boldsymbol{x}, t\right)
    \label{eq:reactant_mass}
\end{equation}
and
\begin{equation}
    m^p\left(t\right) = \sum_{\boldsymbol{x}} \rho^p\left(\boldsymbol{x}, t\right),
    \label{eq:product_mass}
\end{equation}
respectively. Due to mass conservation, we can write the total mixture mass as
\begin{equation}
    m^\mathrm{m} = m^r\left(t\right) + m^p\left(t\right) = \rho_{0} \: \phi \: V.
    \label{eq:total_mass}
\end{equation}
Additionally, we can compute the time-dependent conversion rate
\begin{equation}
    \dot{m}^r\left(t\right) = - \xi \sum_{\boldsymbol{x}} \rho^r\left(\boldsymbol{x}, t\right) r\left(\boldsymbol{x}\right) \text{NN}\left(\boldsymbol{x}\right),
    \label{eq:production_rate}
\end{equation}
which is directly dependent on the form of the reactive boundary condition in \Cref{eq:reactive_boundary_condition_reactant}. Furthermore, we can compare this with a theoretically possible conversion rate in case diffusion were infinitely fast. In that case, whenever reactant fluid is being converted, its density would equilibrate instantly with reactant fluid present at other locations in the simulation box. Consequently, the density at every reactive site is always equal to the average reactant density in the simulation box. Therefore, the maximum theoretically possible conversion rate is
\begin{equation}
    \dot{m}_\mathrm{th}^r\left(t\right) = - \frac{\xi A_r}{\phi V} \: m^r\left(t\right).
    \label{eq:production_rate_theoretical}
\end{equation}
In this equation, the preceding constant is dependent on the geometry and catalyst loading. Combining \Cref{eq:production_rate} and \Cref{eq:production_rate_theoretical}, we can compute a time-dependent effectiveness factor
\begin{equation}
    \alpha\left(t\right) = \frac{\dot{m}^r\left(t\right)}{\dot{m}_\mathrm{th}^r\left(t\right)},
    \label{eq:effectiveness_factor}
\end{equation}
which can be computed directly from the simulations. Moreover, we can combine \Cref{eq:production_rate_theoretical} and \Cref{eq:effectiveness_factor} to obtain the differential equation
\begin{equation}
    \dot{m}^r\left(t\right) = - \frac{\xi A_r}{\phi V} \: \alpha\left(t\right) \: m^r\left(t\right).
    \label{eq:effectiveness_factor_differential}
\end{equation}
To integrate this equation, we assume that the effectiveness factor remains approximately constant during a simulation, $\alpha\left(t\right) = \alpha$. We will later demonstrate that this is a reasonable approximation when the reactant fluid is available and being consumed. Solving \Cref{eq:effectiveness_factor_differential} for the available reactant mass then gives
\begin{equation}
    m^r\left(t\right) = m^\mathrm{m} \: \mathrm{e}^{- \frac{\xi A_r}{\phi V} \alpha t}.
    \label{eq:effectiveness_factor_solution}
\end{equation}
The integration constant is the initial reactant mass, which is equal to the total fluid mass. Given that the geometry is initialized with reactant fluid only, this is also the total mass available for conversion, including the reactant mass that might remain trapped in the geometry.

\subsection{Model validation}\label{sec:model-validation}
\begin{figure}
    \centering
    \includegraphics{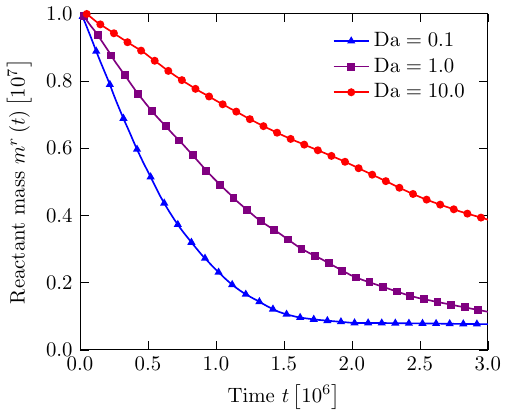}
    \caption{The reactant mass $m^r\left(t\right)$ as a function of the time $t$. Asymptotic decay of the mass fraction, with decay rates dependent on the Damköhler number, can clearly be observed. Besides, it can be noticed that not all the reactant mass is consumed on long time scales. This is due to pockets of reactant fluid remaining trapped in the absence of a nearby reactive site.}
    \label{fig:reactant}
\end{figure}

To validate the correctness of the analytical model, we study the behavior of three simulations of immiscible reactive flows in the bijel-derived geometry for long time scales. We make use of the simulation parameters discussed in \Cref{sec:chemically-reactive-flows}. In \Cref{fig:reactant}, we plot the obtained time evolutions for the total reactant mass using $3,000,000 \Delta t$. As can be seen in this figure, the total reactant mass fraction complies with the predicted exponential decay, with decay rates that are dependent on the Damköhler number. Moreover, it can clearly be spotted that the curve for $\mathrm{Da} = 0.1$ converges to a non-zero value. This indicates that, for this particular simulation setup, regions where the reactant is trapped in pores without reactive sites exist.

When reactant fluid is trapped in pores without reactive sites, product fluid flows around through pores that are not clogged. Important to note here is that the clogging locations and the exact amount of remaining reactant mass are heavily influenced by the numerics and the simulation settings. A slightly wider interface or a small change in the reactive site distribution can have a large effect on the final mass of remaining reactant. To this end, this is not included in the theoretical model.
Clogging in bijel-derived porous structures is physical under certain conditions \cite{Beunen2026BubblesHighlyPorous}. Possible explanations involve slight non-uniformities in the pore structure or effects related to surface roughness. However, this is not something that we can reliably predict with our current simulations and should be studied in a future contribution.

\begin{figure}
    \centering
    \includegraphics{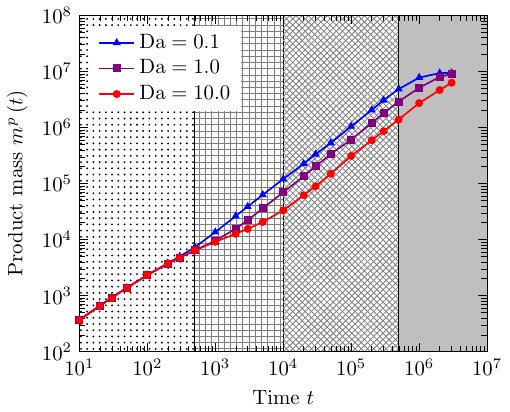}
    \caption{The product mass plotted versus time on a log-log scale. We distinguish four different regions indicated by the gray background patterns. First, we have an initial startup phase with rapid bubble growth. Second, we see the onset of pore-scale effects where the formed bubbles start feeling the influence of the surrounding pore structure. Third, we enter the regime predicted by the analytics in \Cref{eq:effectiveness_factor_solution}, indicated by the linearity of the curve. Finally, we see exhaustion of reactant fluid accompanied by flattening of the product mass curve.}
    \label{fig:production_log}
\end{figure}

In \Cref{fig:production_log}, we plot the evolution of the total product mass on a log-log scale. We distinguish four distinct regimes. First, we have an initial startup phase between $0 \Delta t$ and $500 \Delta t$. This phase is characterized by rapid and highly non-equilibrium bubble growth. The reactant is abundantly available at the reactive sites and is very quickly consumed. Second, between $500 \Delta t$ and $10,000 \Delta t$, we see the onset of pore-scale effects: the formed bubbles start feeling the influence of the surrounding pore structure. This leads to Laplace pressure-related effects; the bubbles tend to become non-spherical, depending on the pore structure. Additionally, as the reactive sites are surrounded by increasing amounts of product fluid, reactant transport to them is slowed down. Third, between $10,000 \Delta t$ and $500,000 \Delta t$, we enter the regime predicted by the analytical equation. In particular, we observe the linear part of the exponential decay of the reactant mass predicted by \Cref{eq:effectiveness_factor_solution}. The linearity of the curves indicates that the total product mass scales as $m^p\left(t\right) \propto t$. This is consistent with the exponential decay predicted by the theoretical model; by Taylor-expanding the decaying exponential in \Cref{eq:effectiveness_factor_solution} around zero, we obtain a dominant term proportional to $t$. Finally, after $500,000 \Delta t$, we enter the regime in which the reactant is being exhausted, and we see a flattening of the curve. Fully reaching the plateau for larger $\mathrm{Da}$ turned out to be infeasible due to the immense computing resources required (note the log-scale of the time axis).

\begin{figure*}
    \centering
    \begin{tabular}{ccc}
    \begin{subfigure}[b]{.3\linewidth}
        \centering
        \includegraphics[width=\linewidth, trim=50 50 50 100, clip]{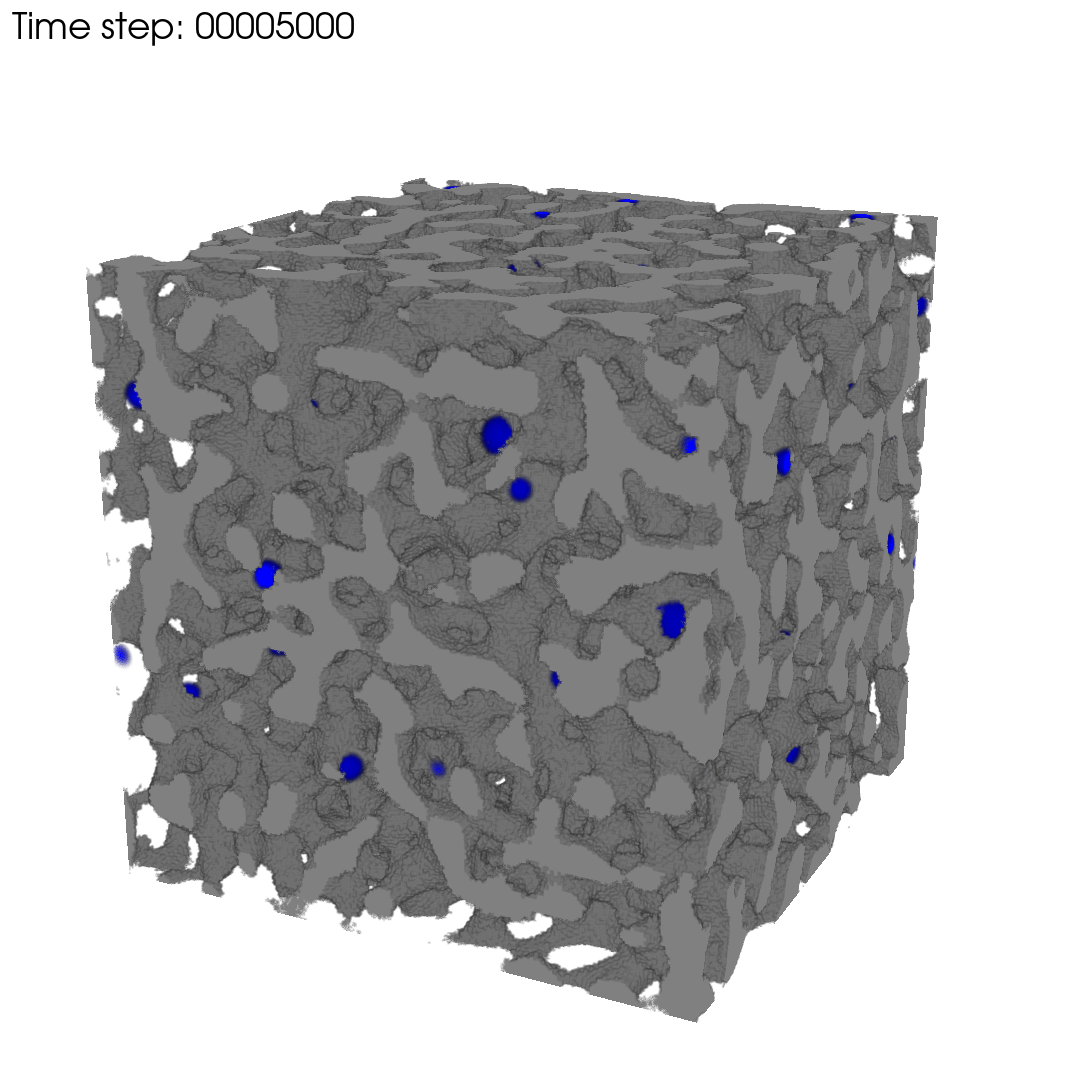}
        \caption{$5,000 \Delta t$}
    \end{subfigure}
    &
    \begin{subfigure}[b]{.3\linewidth}
        \centering
        \includegraphics[width=\linewidth, trim=50 50 50 100, clip]{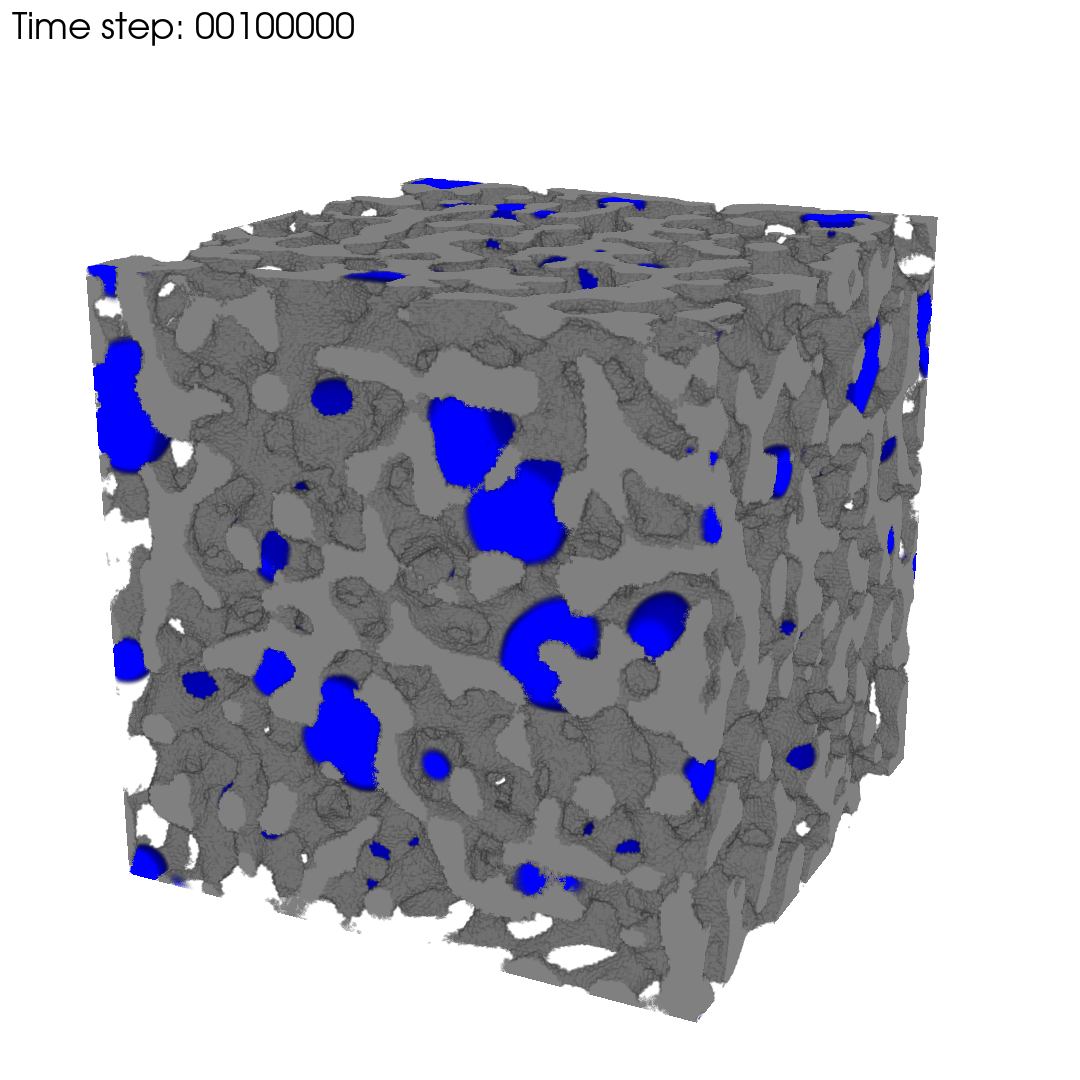}
        \caption{$100,000 \Delta t$}
    \end{subfigure}
    &
    \begin{subfigure}[b]{.3\linewidth}
        \centering
        \includegraphics[width=\linewidth, trim=50 50 50 100, clip]{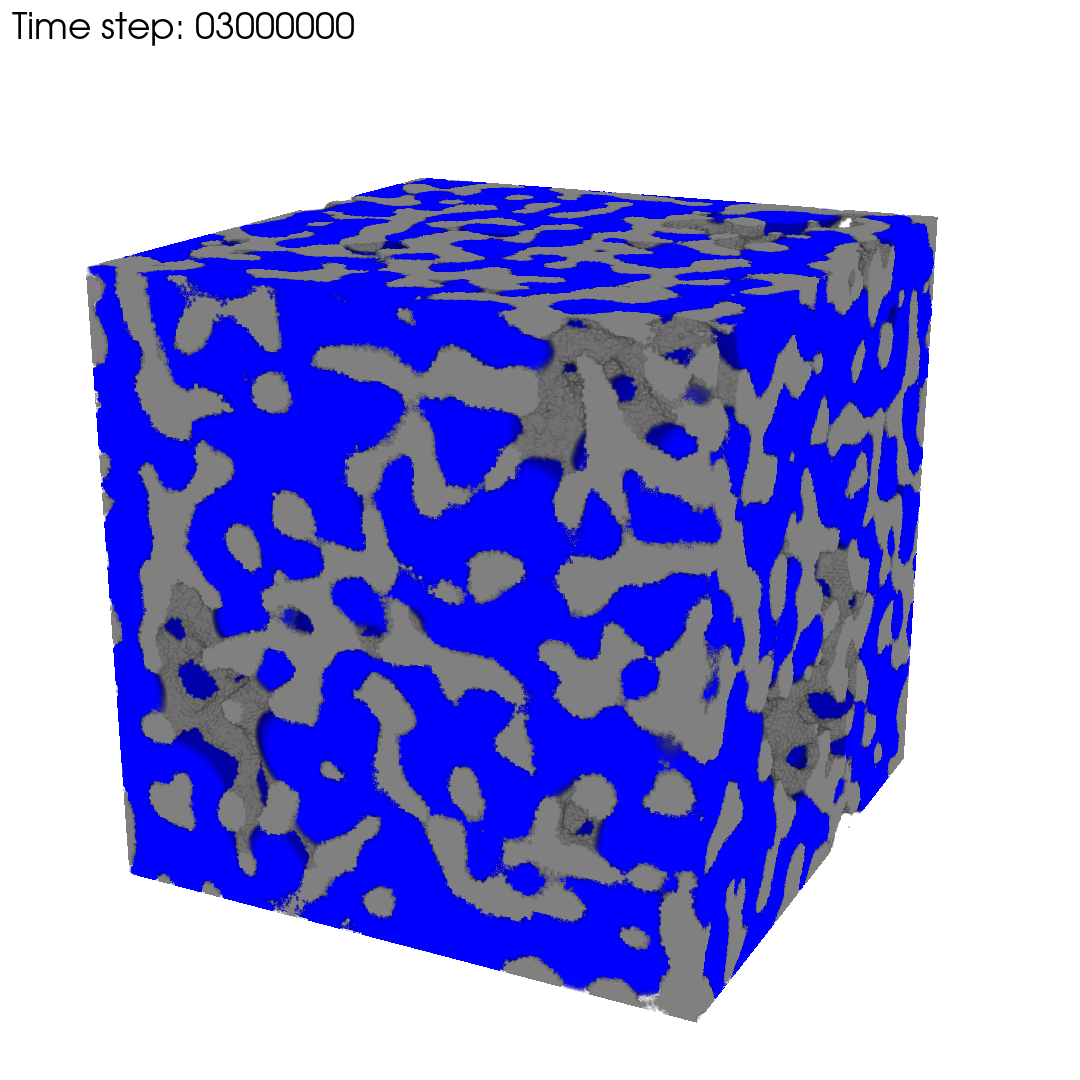}
        \caption{$3,000,000 \Delta t$}
    \end{subfigure}
    \end{tabular}
    \caption{Snapshots of the simulation with $\mathrm{Da} = 1.0$ for three simulation regimes. The pore geometry is rendered in gray, with the reactant fluid being made transparent and the product fluid shown in blue. Figure \textbf{(a)} shows a snapshot of the simulation at $5,000 \Delta t$, when the bubbles have grown large enough to start experiencing effects from the pore structure around them. Image \textbf{(b)}, at $100,000 \Delta t$, is a snapshot of the simulation after having entered the linear regime predicted by the analytics. In image \textbf{(c)}, at $3,000,000 \Delta t$, exhaustion of the reactant fluid for long time scales can be observed.}
    \label{fig:snapshots}
\end{figure*}

To further illustrate the regimes shown in \Cref{fig:production_log}, snapshots of the simulation with $\mathrm{Da} = 1.0$ are shown in \Cref{fig:snapshots}. In this figure, the left snapshot at $5,000 \Delta t$ shows the regime where pore-scale effects begin to appear. Bubbles are growing in the pores and are just beginning to experience the influence of their surrounding pore structure. The snapshot in the center at $100,000 \Delta t$ illustrates the simulation state in the regime where the reactant mass decay can be predicted by \Cref{eq:effectiveness_factor_solution}. Finally, the right snapshot at $3,000,000 \Delta t$ visualizes the simulation state where the reactant fluid is almost exhausted. Only pockets of reactant fluid, trapped in the product fluid and unable to reach the reactive sites, remain.

\begin{figure}
    \centering
    \includegraphics{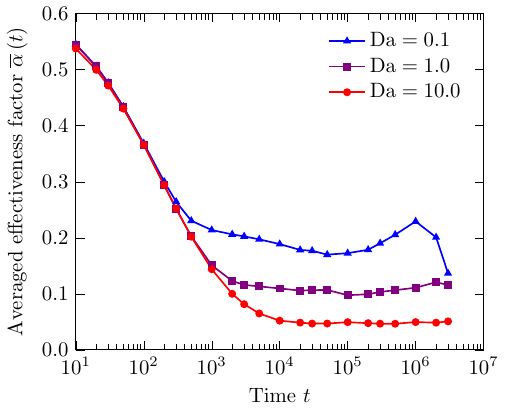}
    \caption{Evolution of the average of the effectiveness factor, computed for the preceding time steps. At the beginning of the simulation, the value for the effectiveness factor is artificially high due to the reactant being very abundantly available at the reactive sites. However, we observe that after bubbles have been formed in the pores around $10,000 \Delta t$, the averaged effectiveness factor remains approximately constant for the time that the reactant remains abundantly available. This is consistent with the constant effectiveness factor assumption, used to be able to integrate \Cref{eq:effectiveness_factor_differential}.}
    \label{fig:effectiveness_factor_avg_log}
\end{figure}

In \Cref{fig:effectiveness_factor_avg_log}, the evolution of the averaged effectiveness factor is plotted, where the averaging happens with equal weighting over the preceding time steps,
\begin{equation}
    \overline{\alpha}\left(t\right) = \frac{1}{t} \sum_{i=0}^{t} \alpha\left(i\right).
    \label{eq:effectiveness_factor_avg}
\end{equation}
At the beginning of the simulation, the value for the averaged effectiveness factor is very high due to the reactant fluid being abundantly available at the reactive sites. However, we observe that after approximately $10,000 \Delta t$, when bubbles have been formed inside the pores, the values remain close to constant. This remains the case for several orders of magnitude of $\Delta t$ and is consistent with the assumption of a constant effectiveness factor in \Cref{eq:effectiveness_factor_differential}. However, when the reactant fluid is nearly exhausted, we observe an increase in the averaged effectiveness factor, followed by a drop to zero. This increase in the effectiveness factor is related to coalescence dynamics when the geometry is almost filled with product fluid. The coalescence of many bubbles in rapid succession violently moves the remaining reactant fluid, resulting in faster conversion.

\begin{figure}
    \centering
    \includegraphics{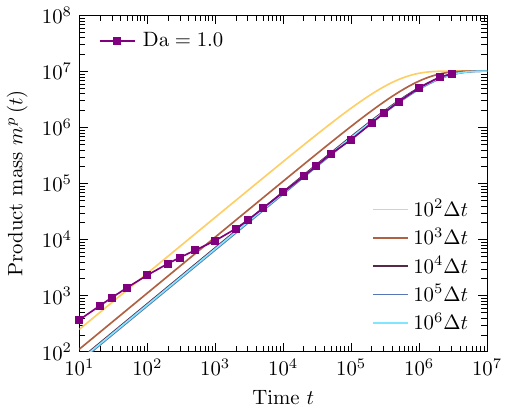}
    \caption{Fitting accuracy of the effectiveness factor computed from preceding time steps using \Cref{eq:effectiveness_factor_solution}. The violet curve with square marks shows the data obtained from the simulation with $\mathrm{Da} = 1.0$. The colored lines denote the curves for \Cref{eq:effectiveness_factor_solution} that can be obtained by fitting the evolution of the reactant mass on the displayed orders of magnitude of $\Delta t$. It can be observed that $10,000 \Delta t$ is already enough for the curves to overlap and yield realistic predictions.}
    \label{fig:production_fit_log}
\end{figure}

Finally, we analyse the fitting accuracy of \Cref{eq:effectiveness_factor_solution} on the reactant and product masses. In Figure \Cref{fig:production_fit_log}, the evolution of the product mass is plotted for the simulation with $\mathrm{Da} = 1.0$ on a log-log scale. We fit \Cref{eq:effectiveness_factor_solution} to the simulation data using a variety of time steps $\Delta t$. This enables a comparison of the amount of simulation data required to accurately predict the mass flow from reactant to product in the simulations. As can be seen in \Cref{fig:production_fit_log}, the curve predicted by the fit starts to overlap with the simulation data when using about $10,000 \Delta t$ to compute the effectiveness factor. When increasing this number further, we observe that excellent fitting accuracy can be obtained with as little as $50,000 \Delta t$, with predictions being at most three percent off from the $\overline{\alpha}\left(t\right)$ values. Given that this reduction in the required number of time steps greatly decreases the computational cost of a simulation, it also greatly increases the number of simulations that we are able to run. To this end, we are able to perform parameter searches to optimize the wetting properties of the geometry.

\subsection{Contact angle optimization}\label{sec:contact-angle-optimization}
\begin{figure}[!hb]
    \centering
    \includegraphics{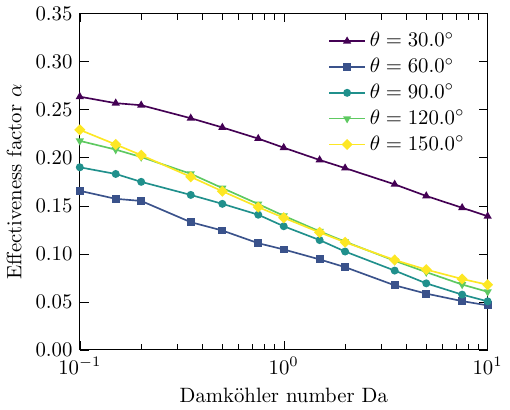}
    \caption{The effectiveness factor $\alpha$ fitted on the first $50,000 \Delta t$ of a simulation plotted against the Damköhler number $\mathrm{Da}$. As expected, we observe a decrease in the effectiveness factor across all contact angles as the Damköhler number increases. Furthermore, we see the effectiveness factor increasing significantly for the superhydrophilic contact angle ($\theta = 30.0^\circ$) for all Damköhler numbers.}
    \label{fig:wetting_effectiveness_factor1}
\end{figure}

As discussed in the previous section, we decrease the number of time steps from $3,000,000 \Delta t$ to $50,000 \Delta t$ to enable parameter searches on optimal wetting properties for bijel-derived porous structures with reasonable computational resources. This enables us to run a large number of simulations to study the influence of the contact angle on the effectiveness factor. We fit the evolution of the reactant mass with \Cref{eq:effectiveness_factor_solution} to obtain a prediction for the effectiveness factor that also holds for long time scales. In \Cref{fig:wetting_effectiveness_factor1}, these predictions for the effectiveness factor are plotted against the Damköhler number for five different contact angles. As can be seen, increasing the Damköhler number monotonically decreases the effectiveness factor for all contact angles. Additionally, it can be observed that decreasing the contact angle from $150^\circ$ (superhydrophobic) to $60^\circ$ (hydrophilic) decreases the effectiveness factor as well. However, this behavior shifts when the contact angle is decreased further to $30^\circ$ (superhydrophilic); the effectiveness factor shoots up to a value that is higher than what can be achieved with a $150^\circ$ contact angle.

\begin{figure}
    \centering
    \includegraphics{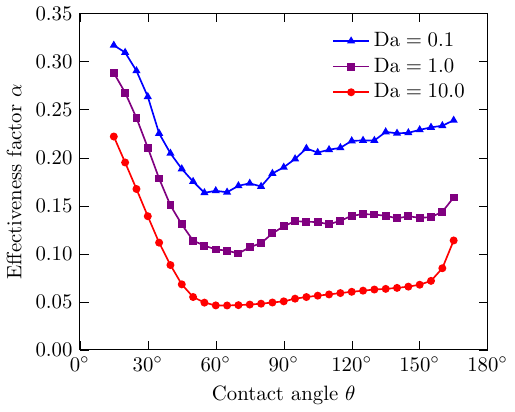}
    \caption{The effectiveness factor $\alpha$ fitted on the first $50,000 \Delta t$ of a simulation plotted against the contact angle $\theta$. We show simulation data for the same Damköhler numbers as in \Cref{sec:model-validation}. It can be seen that the effectiveness factor tends to be lower for moderate contact angles. However, for superhydrophilic and superhydrophobic contact angles, the effectiveness factor tends to increase significantly.}
    \label{fig:wetting_effectiveness_factor2}
\end{figure}

We further investigate this effect for three different Damköhler numbers. The resulting behavior of the effectiveness factor versus the contact angle is plotted in \Cref{fig:wetting_effectiveness_factor2}. For all the Damköhler numbers, the effectiveness factor peaks for superhydrophilic and superhydrophobic substrates. Additionally, all the curves indicate a minimum in the effectiveness factor for hydrophilic contact angles between $55^\circ$ and $70^\circ$. While making the substrate more hydrophobic from this point on does not massively increase the effectiveness factor, much larger gains can be realized when the substrate is made hydrophilic. Especially for the $\mathrm{Da} = 10.0$ case, the increase is very significant. Decreasing the contact angle from $65^\circ$ to $15^\circ$ increases the effectiveness factor almost fivefold. For the other Damköhler numbers, we see a similar effect: for $\mathrm{Da} = 1.0$ the increase is almost threefold, and for $\mathrm{Da} = 0.1$ almost twofold.

This drastic increase in the effectiveness factor for superhydrophilic contact angles is due to an improvement in bubble departure speed, subsequently leading to a higher availability of reactant mass at the reactive sites. Several effects contribute to this enhanced bubble departure. The surface area between a bubble and the porous structure is generally smaller for smaller contact angles, which reduces the detachment energy. Additionally, when a bubble does not detach before it grows to the size of the surrounding pore, effects related to the Laplace pressure tend to be more prominent than for moderate contact angles. Due to the low contact angle, small non-uniformities in the pore structure can lead to large Laplace pressure differences. This tends to move a bubble away from a reactive site, freeing up space for more reactant to flow to a reactive site. Finally, while it is hard to directly observe in the simulations due to their size, we expect that the surface roughness leads to additional effects related to contact angle pinning.

\section{Conclusion}\label{sec:conclusion}
As identified by previous work, these kinds of porous structures have great potential for catalysis applications due to their particular geometrical properties. In this work, we investigated the influence of the Damköhler number and wettability properties of bijel-derived porous structures for immiscible reactive flows. We presented a simplified analytical model that allows us to predict the effectiveness factor of a structure on long-time scales using only a small number of simulation time steps. Hereafter, we analyzed a small number of reactive flow simulations for long durations, all the way up to the point where reactant exhaustion starts playing a role. These simulations confirmed the validity and accuracy of our analytical model. Finally, using our analytical model and fitting it to short simulations, we studied the influence of the wetting behavior on bijel-derived porous structures. We found that both very small and very large contact angles, i.e., superhydrophilic and superhydrophobic structures, can drastically enhance the effectiveness factor. In particular, using a superhydrophilic porous structure yields effectiveness factors that are several times higher due to an enhancement in bubble departure speed.

\begin{acknowledgments}
    We acknowledge financial support from the Bavarian Ministry for Economy, State Development and Energy -- Project H2Season (Grant number RMF-SG20-3410-6-10-11). Additionally, we thank the Gauss Centre for Supercomputing e.V. (\url{www.gauss-centre.eu}) for funding this project by providing computing time through the John von Neumann Institute for Computing (NIC) on the GCS Supercomputer JUWELS at the Jülich Supercomputing Centre (JSC).
\end{acknowledgments}

\section*{Data availability}
The data that support the findings of this study are openly available at \url{https://doi.org/10.5281/zenodo.XXXXXXXX}.

\appendix

\section{Generalized central moments collision operator}\label{sec:generalized-central-moments-collision-operator}
To achieve numerical stability improvements beyond what the standard BGK collision operator offers, collision operators with multiple relaxation times are often used. These kinds of collision operators generally transform the populations into raw moments, relax these moments with different relaxation times, and transform them back into populations. While the increased numerical stability of using such procedures is non-negligible, problems still arise for low kinematic viscosities. To achieve further stability gains, several authors suggested not just relaxing raw moments of the populations, but also transforming these moments into a reference frame that is comoving with the populations. Applying the collision step to so-called central moments (CMs) results in drastic stability gains in the low kinematic viscosity limit \cite{DeRosis2020MultiphysicsFlowSimulations,Coreixas2019ComprehensiveComparisonCollision, Coreixas2020ImpactCollisionModels}.

Compared with collision operators that use raw moments, collision operators that use central moments differ in the sense that the relaxation scheme is dependent on the velocity vector at a particular lattice site. Accordingly, the transformation matrix from populations to moments $\mathbf{T}$ is different at every lattice position $\boldsymbol{x}$ and time $t$. To perform the collision step, we follow the procedure outlined by De Rosis and Coreixas \cite{DeRosis2020MultiphysicsFlowSimulations}. However, we generalize this procedure to arbitrary equilibria while relaxing only the five central moments associated with the viscous shear stresses.

To demonstrate the procedure, we first write the central moments collision operator in \Cref{eq:single_phase_collision} in vector notation
\begin{equation}
    \boldsymbol{\Omega}^{\mathrm{CM}} = \boldsymbol{f}^{\mathrm{eq}} - \boldsymbol{f} + \mathbf{T}^{-1} \, \left(\mathbf{I} - \mathbf{R}\right) \, \mathbf{T} \boldsymbol{f}.
    \label{eq:collision_operator}
\end{equation}
Here, $\boldsymbol{f} = \left[f_0, \ldots, f_i, \ldots, f_{18}\right]^{\top}$ and $\boldsymbol{f}^{\mathrm{eq}} = \left[f_0^{\mathrm{eq}}, \ldots, f_i^{\mathrm{eq}}, \ldots, f_{18}^{\mathrm{eq}}\right]^{\top}$ are column vectors of the populations and equilibrium populations, respectively. Additionally, we have the transformation matrix $\mathbf{T} = \left[\boldsymbol{t}_0, \ldots, \boldsymbol{t}_i, \ldots, \boldsymbol{t}_{18},\right]$ from populations to central moments, with vectors
\begin{equation}
    \boldsymbol{t}_i = \left(
    \begin{array}{c}
        1                                                               \\
        \overline{c}_{ix}                                               \\
        \overline{c}_{iy}                                               \\
        \overline{c}_{iz}                                               \\
        \overline{c}_{ix}^2 + \overline{c}_{iy}^2 + \overline{c}_{iz}^2 \\
        \overline{c}_{ix}^2 - \overline{c}_{iy}^2                       \\
        \overline{c}_{iy}^2 - \overline{c}_{iz}^2                       \\
        \overline{c}_{ix} \overline{c}_{iy}                             \\
        \overline{c}_{ix} \overline{c}_{iz}                             \\
        \overline{c}_{iy} \overline{c}_{iz}                             \\
        \overline{c}_{ix}^2 \overline{c}_{iy}                           \\
        \overline{c}_{ix} \overline{c}_{iy}^2                           \\
        \overline{c}_{ix}^2 \overline{c}_{iz}                           \\
        \overline{c}_{ix} \overline{c}_{iz}^2                           \\
        \overline{c}_{iy}^2 \overline{c}_{iz}                           \\
        \overline{c}_{iy} \overline{c}_{iz}^2                           \\
        \overline{c}_{ix}^2 \overline{c}_{iy}^2                         \\
        \overline{c}_{ix}^2 \overline{c}_{iz}^2                         \\
        \overline{c}_{iy}^2 \overline{c}_{iz}^2
    \end{array}
    \right)
    \label{eq:forward_transformation_vector}
\end{equation}
The entries of the transformation matrix are computed based on shifting the lattice vectors with the local velocity vector
\begin{equation}
    \overline{\boldsymbol{c}}_i = \boldsymbol{c}_i - \boldsymbol{u},
    \label{eq:local_fluid_velocity}
\end{equation}
where the indices $x$, $y$, and $z$ are used to index the Cartesian directions of these vectors. Finally, we have the identity matrix $\mathbf{I}$ and the relaxation matrix
\begin{equation}
    \mathbf{R} = \operatorname{diag}\left(1, 1, 1, 1, 1, \frac{1}{\tau}, \frac{1}{\tau}, \frac{1}{\tau}, \frac{1}{\tau}, \frac{1}{\tau}, 1, \ldots, 1\right).
    \label{eq:relaxation_matrix}
\end{equation}
Contrary to the approach proposed by De Rosis and Coreixas \cite{DeRosis2020MultiphysicsFlowSimulations}, we never use the equilibrium populations to compute central moments in \Cref{eq:collision_operator}. Instead, due to the formulation of the collision operator in this equation, we compute central moments directly from the full populations. Given the structure of the matrices, this leads to non-zero values for only five of the nineteen post-collision central moments
\begin{equation}
    \boldsymbol{k}^\star = \left(\mathbf{I} - \mathbf{R}\right)\mathbf{T} \boldsymbol{f}.
    \label{eq:post_collision_moments_vector}
\end{equation}
These non-zero post-collision central moments can be found at indices $i = 5 \ldots 9$, and are computed as
\begin{align}
    k_5^\star &= \left(1 - \frac{1}{\tau}\right) \sum_i f_i \left(\overline{c}_{ix}^2 - \overline{c}_{iy}^2\right), \\
    k_6^\star &= \left(1 - \frac{1}{\tau}\right) \sum_i f_i \left(\overline{c}_{iy}^2 - \overline{c}_{iz}^2\right), \\
    k_7^\star &= \left(1 - \frac{1}{\tau}\right) \sum_i f_i \overline{c}_{ix} \overline{c}_{iy}, \\
    k_8^\star &= \left(1 - \frac{1}{\tau}\right) \sum_i f_i \overline{c}_{ix} \overline{c}_{iz}, \\
    k_9^\star &= \left(1 - \frac{1}{\tau}\right) \sum_i f_i \overline{c}_{iy} \overline{c}_{iz}.
    \label{eq:post_collision_moments}
\end{align}
As a result, we only need to transform five central moments back into population space to compute the post-collision populations. This results in a significantly simpler inverse transformation, and hence in a simpler and computationally more efficient numerical scheme. The expressions computed for $\mathbf{T}^{-1} \boldsymbol{k}^\star$ are given in \Cref{eq:back_transformation_vector} and can be implemented directly in a numerical solver.
\onecolumngrid
\begin{equation}
    \mathbf{T}^{-1} \boldsymbol{k}^\star = \left(
    \begin{array}{rrr}
        \frac{1}{3} \left(u_x^2+u_x-u_y^2-u_z^2+1\right) k_5^\star   & +\frac{1}{6} \left(u_x^2+u_x-u_y^2-u_z^2+1\right) k_6^\star & -(2 u_x+1) u_y k_7^\star-(2 u_x+1) u_z k_8^\star \\
        \frac{1}{3} \left(u_x^2-u_x-u_y^2-u_z^2+1\right) k_5^\star   & +\frac{1}{6} \left(u_x^2-u_x-u_y^2-u_z^2+1\right) k_6^\star & -(2 u_x-1) u_y k_7^\star-(2 u_x-1) u_z k_8^\star \\
        \frac{1}{6} \left(u_x^2-u_y^2+u_z^2-u_y-1\right) k_5^\star   & +\frac{1}{6} \left(-u_x^2+u_y^2-u_z^2+u_y+1\right) k_6^\star & -(2 u_y+1) u_x k_7^\star-(2 u_y+1) u_z k_9^\star \\
        \frac{1}{6} \left(u_x^2-u_y^2+u_z^2+u_y-1\right) k_5^\star   & +\frac{1}{6} \left(-u_x^2+u_y^2-u_z^2-u_y+1\right) k_6^\star & -(2 u_y-1) u_x k_7^\star-(2 u_y-1) u_z k_9^\star \\
        \frac{1}{6} \left(u_x^2+u_y^2-u_z^2-u_z-1\right) k_5^\star   & +\frac{1}{3} \left(u_x^2+u_y^2-u_z^2-u_z-1\right) k_6^\star & -(2 u_z+1) u_x k_8^\star-(2 u_z+1) u_y k_9^\star \\
        \frac{1}{6} \left(u_x^2+u_y^2-u_z^2+u_z-1\right) k_5^\star   & +\frac{1}{3} \left(u_x^2+u_y^2-u_z^2+u_z-1\right) k_6^\star & -(2 u_z-1) u_x k_8^\star-(2 u_z-1) u_y k_9^\star \\
        \frac{1}{12} \left(-u_x^2-u_x+2u_y^2+2u_y\right) k_5^\star   & +\frac{1}{12} \left(u_x^2+u_x+u_y^2+u_y\right) k_6^\star & +\frac{1}{12} (12 u_y u_x+6 u_x+6 u_y+3) k_7^\star \\
        \frac{1}{12} \left(-u_x^2-u_x+2u_y^2-2u_y\right) k_5^\star   & +\frac{1}{12} \left(u_x^2+u_x+u_y^2-u_y\right) k_6^\star & +\frac{1}{12} (12 u_y u_x-6 u_x+6 u_y-3) k_7^\star \\
        \frac{1}{12} \left(-u_x^2-u_x+2u_z^2+2u_z\right) k_5^\star   & +\frac{1}{12} \left(-2 u_x^2-2u_x+u_z^2+u_z\right) k_6^\star & +\frac{1}{12} (12 u_z u_x+6 u_x+6 u_z+3) k_8^\star \\
        \frac{1}{12} \left(-u_x^2-u_x+2u_z^2-2u_z\right) k_5^\star   & +\frac{1}{12} \left(-2 u_x^2-2u_x+u_z^2-u_z\right) k_6^\star & +\frac{1}{12} (12 u_z u_x-6 u_x+6 u_z-3) k_8^\star \\
        \frac{1}{12} \left(-u_x^2+u_x+2u_y^2+2u_y\right) k_5^\star   & +\frac{1}{12} \left(u_x^2-u_x+u_y^2+u_y\right) k_6^\star & +\frac{1}{12} (12 u_y u_x+6 u_x-6 u_y-3) k_7^\star \\
        \frac{1}{12} \left(-u_x^2+u_x+2u_y^2-2u_y\right) k_5^\star   & +\frac{1}{12} \left(u_x^2-u_x+u_y^2-u_y\right) k_6^\star & +\frac{1}{12} (12 u_yu_x-6 u_x-6 u_y+3) k_7^\star \\
        \frac{1}{12} \left(-u_x^2+u_x+2 u_z^2+2 u_z\right)k_5^\star  & +\frac{1}{12} \left(-2 u_x^2+2 u_x+u_z^2+u_z\right) k_6^\star & +\frac{1}{12} (12 u_z u_x+6u_x-6 u_z-3) k_8^\star \\
        \frac{1}{12} \left(-u_x^2+u_x+2 u_z^2-2 u_z\right) k_5^\star & +\frac{1}{12} \left(-2u_x^2+2 u_x+u_z^2-u_z\right) k_6^\star & +\frac{1}{12} (12 u_z u_x-6 u_x-6 u_z+3)k_8^\star \\
        \frac{1}{12} \left(-u_y^2-u_y-u_z^2-u_z\right) k_5^\star     & +\frac{1}{12} \left(-2 u_y^2-2u_y+u_z^2+u_z\right) k_6^\star & +\frac{1}{12} (12 u_z u_y+6 u_y+6 u_z+3) k_9^\star \\
        \frac{1}{12}\left(-u_y^2-u_y-u_z^2+u_z\right) k_5^\star      & +\frac{1}{12} \left(-2 u_y^2-2u_y+u_z^2-u_z\right) k_6^\star & +\frac{1}{12} (12 u_z u_y-6 u_y+6 u_z-3) k_9^\star \\
        \frac{1}{12}\left(-u_y^2+u_y-u_z^2-u_z\right) k_5^\star      & +\frac{1}{12} \left(-2 u_y^2+2u_y+u_z^2+u_z\right) k_6^\star & +\frac{1}{12} (12 u_z u_y+6 u_y-6 u_z-3) k_9^\star \\
        \frac{1}{12}\left(-u_y^2+u_y-u_z^2+u_z\right) k_5^\star      & +\frac{1}{12} \left(-2 u_y^2+2u_y+u_z^2-u_z\right) k_6^\star & +\frac{1}{12} (12 u_z u_y-6 u_y-6 u_z+3) k_9^\star \\
        \frac{1}{3}\left(-2 u_x^2+u_y^2+u_z^2\right) k_5^\star       & +\frac{1}{3} \left(-u_x^2-u_y^2+2 u_z^2\right) k_6^\star & +4u_x u_y k_7^\star+4 u_x u_z k_8^\star+4 u_y u_z k_9^\star
    \end{array}
    \right)
    \label{eq:back_transformation_vector}
\end{equation}
\twocolumngrid
Finally, a few important points need to be noted regarding the proposed formulation of the central moments collision operator. First, \Cref{eq:back_transformation_vector} can be implemented with even higher computational efficiency by precomputing several common subexpressions.
Second, given that the equilibrium populations are not passed through the matrices, and only central moments related to viscous shear stresses are being relaxed, this formulation is generic and works for a whole class of equilibria. In particular, any equilibrium modelling the Boltzmann distribution can be used, independent of the truncation order. Furthermore, unlike previous formulations, this formulation of the central moments collision operator can also be straightforwardly applied in simulations with enhanced color-gradient equilibria that make use of non-local density gradients \cite{Leclaire2013EnhancedEquilibriumDistribution, Leclaire2017GeneralizedThreedimensionalLattice}.
Third, in contrast to the formulation used by De Rosis and Coreixas \cite{DeRosis2020MultiphysicsFlowSimulations}, body forces cannot be implemented directly using the popular model by Guo \textit{et al.} \cite{Guo2002DiscreteLatticeEffects}. This forcing model is dependent on the exact collision operator being used, and in this case requires information on more than just five central moments. We can elegantly resolve this problem by opting for a forcing model that is independent of the central moments collision operator and depends only on the form of the equilibrium distribution: the exact difference scheme by Kupershtokh \textit{et al.} \cite{Kupershtokh2009EquationsStateLattice}, as written in \Cref{eq:operator_perturbation}. Given that the formulation of this forcing model is also independent of the exact form of the equilibrium, this does not add any additional limitations on the form of the equilibrium as discussed under the second point.


%
\end{document}